\documentclass[a4paper,11pt]{article}
\usepackage{pos}

\usepackage{mathrsfs}
\usepackage{amsmath}
\usepackage{graphicx}
\usepackage{amsfonts,amsbsy}
\usepackage{nicefrac}
\usepackage{xcolor}
\usepackage{comment}

\definecolor{lcolor}{rgb}{0.5,0,0}
\definecolor{citcolor}{rgb}{0,0.3,0.0}
\definecolor{ao(english)}{rgb}{0.0, 0.5, 0.0}

\definecolor{applegreen}{rgb}{0.55, 0.71, 0.0}
\definecolor{cadetblue}{rgb}{0.37, 0.62, 0.63}
\definecolor{cadet}{rgb}{0.33, 0.41, 0.47}
\definecolor{byzantine}{rgb}{0.74, 0.2, 0.64}
\definecolor{orange}{rgb}{1.0, 0.5, 0.0}

\def\bea{\begin{eqnarray}}
\def\eea{\end{eqnarray}}

\def\be{\begin{equation}}
\def\ee{\end{equation}}

\newcommand{\Q}{Q_s}
\newcommand{\pdv}[2][]{\frac{\partial{#1}}{\partial{#2}}}
\newcommand{\vb}{\vec}
\newcommand{\ud}{\mathrm{d}}

\newcommand{\fig}{Fig.~}

\newcommand{\se}{Sec.~}

\newcommand{\pToFigs}{.}

\title{Early time dynamics and constraints on medium evolution}

\author*[\dag]{Kirill Boguslavski}
\affiliation{Institute for Theoretical Physics, Technische Universit\"{a}t Wien, 1040 Vienna, Austria}
\emailAdd{kirill.boguslavski@tuwien.ac.at}

\notes{\note{The author is grateful to C.~Andres, J.~Berges, T.~Butler, L.~de Bruin, S.K.~Das, X.~Du, P.~Hotzy, A.~Kurkela, T.~Lappi, F.~Lindenbauer, D.I.~M\"uller, J.~Pawlowski, J.~Peuron, A.V.~Sadofyev, and S.~Schlichting for valuable discussions and collaboration on some of the work presented here. He would also like to thank the organizers for the exciting conference.
  This research was funded in whole, or in part, by the Austrian Science Fund (FWF) project P 34455-N. 
}}

\abstract{In relativistic heavy-ion collisions, a quark-gluon plasma (QGP) is created whose pre-equilibrium evolution includes a rich variety of exciting phenomena of Quantum Chromodynamics. In these Proceedings, we provide a short overview of our current understanding of the pre-equilibrium dynamics of the QGP from a weak-coupling perspective and highlight recent developments. Focusing on new studies, we then discuss how initial stages of the QGP medium may impact hard probes including direct photons, dileptons, jets, and heavy quarks. This introduces promising opportunities to obtain experimental signatures of the pre-equilibrium QGP.
}

\FullConference{HardProbes2023\\
 26-31 March 2023\\
 Aschaffenburg, Germany\\}

\begin{document}
\maketitle


\section{Introduction}

In high-energy heavy-ion collision experiments, a quark-gluon plasma (QGP) is created shortly after the collision and undergoes different phases as it cools down during its evolution. Starting with a pre-equilibrium regime, it becomes eventually describable by viscous hydrodynamics before hadronizing and yielding detector signals. Among others, prominent signatures of the QGP involve jet quenching, heavy quark diffusion, quarkonium suppression, or enhanced spectra \cite{Apolinario:2022vzg, Geurts:2022xmk}. 

In these Proceedings, we will provide an overview of our current understanding of the initial stages in relativistic heavy-ion collisions and on what we can learn about the pre-equilibrium properties of Quantum Chromodynamics (QCD) and of the QGP in particular. While this is a large and active area of research, we will discuss several recent developments to initial stages and focus on the opportunities that hard probes provide to study features of the initial stages experimentally.

In particular, the structure of this work follows two research questions. 
In \se \ref{sec:init_stages} we will address the question: `What are the initial stages of the QGP?' There has been significant progress in QCD calculations over the past decades that employ an interplay of different methods and models. Their aim is to understand the non-equilibrium properties of QCD and to describe the pre-equilibrium evolution of the QGP in heavy-ion collisions. 
In \se \ref{sec:probe_pre_equ} we ask about experimental traces of the pre-equilibrium dynamics, addressing the questions: `How can we probe them experimentally? What are their signatures?' We will discuss new developments, predictions, and opportunities concerning the medium properties of the pre-equilibrium QGP, how these properties affect hard probes, and what we can learn from them. 
Special focus will be given to the jet quenching parameter $\hat q$ and the heavy-quark diffusion coefficient $\kappa$, which are transport coefficients that enter phenomenological evolution equations of jets, heavy quarks, and quarkonia. Finally, we will conclude in \se \ref{sec:concl}.


\section{Initial stages of the Quark-Gluon plasma}
\label{sec:init_stages}

\begin{figure}[t]
    \centering
    \includegraphics[width=\textwidth]{\pToFigs/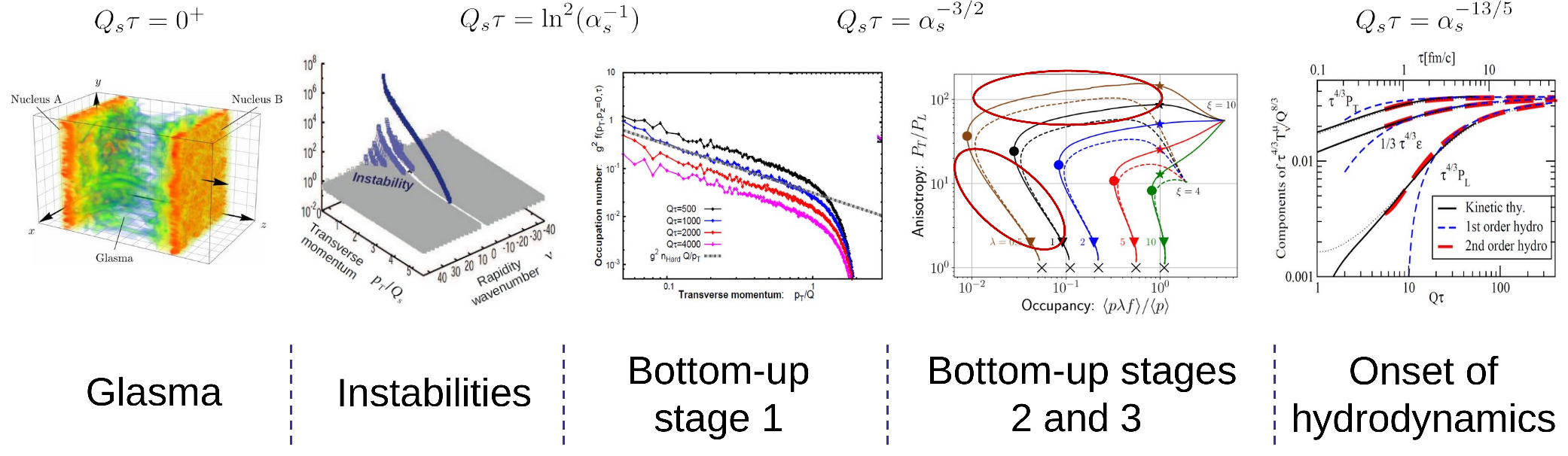}
    \caption{Overview sketch of different phases of the pre-equilibrium evolution of the QGP from a weak-coupling perspective. Figures are taken from (left to right) \cite{Ipp:2017lho, Berges:2014yta, Berges:2013eia, Boguslavski:2023alu, Kurkela:2015qoa}, the 4.~panel is based on a figure from \cite{Kurkela:2015qoa}.}
    \label{fig:init_stages}
\end{figure}

\subsection{Initial stages in heavy-ion collisions -- current picture}
\label{sec:init_stages_overview}

We will start with a short overview of how the QGP evolves during the initial pre-equilibrium stages (see \cite{Schlichting:2019abc, Berges:2020fwq} for recent reviews). The stages are summarized pictorially in \fig \ref{fig:init_stages}. This picture is based on our current understanding according to weak-coupling methods, which are classical-statistical real-time simulations for the first three panels (as long as occupancies are large $f \gg 1$) and QCD kinetic theory for the last three panels (once quasiparticles have formed). Hence, the `bottom-up stage 1' (see below) in the third panel can be studied with both approaches. The last panel overlaps with hydrodynamic simulations. We will provide more details in the following.

\subsubsection*{Strong initial fields: classical-statistical lattice simulations}

Shortly after the collision at (proper) time $\Q \tau = 0^+$, strong chromo-electromagnetic fields are generated between the colliding nuclei that act as sources. The resulting initial state is called `Glasma' \cite{Gelis:2012ri}. Its fields are nonperturbatively large and point initially into the longitudinal (beam) direction 
$\langle \text{Tr}(B_\perp B_\perp) \rangle \sim \langle \text{Tr}(E_\perp E_\perp) \rangle \ll \langle \text{Tr}(B_z B_z) \rangle \sim \langle \text{Tr}(E_z E_z) \rangle \sim 1/\alpha_s$, 
with the coupling constant $\alpha_s(\Q) = g^2/4\pi \ll 1$. The saturation scale $\Q$ is the characteristic transverse momentum scale $\langle p_\perp^2 \rangle \sim \Q^2 \gg \Lambda_{QCD}^2$ of the system. In contrast, the longitudinal momentum $\langle p_z^2 \rangle \approx 0$ vanishes initially due to approximate boost invariance in central collisions. However, boost invariance is broken due to vacuum fluctuations. This leads to plasma instabilities that occupy momentum modes with finite $p_z \approx \nu/\tau$, as visualized in the second panel of \fig \ref{fig:init_stages} (see, e.g., \cite{Berges:2014yta} and references therein). 

The system's dynamics can be described by classical equations of motion, which lead to a gluon-dominated over-occupied plasma at $\Q \tau \sim 1$ up to logarithms in $\alpha_s$, i.e., around $\tau \sim 0.1 - 0.2$ fm/c. In this regime, the large gluon distribution is related to classical-statistical fields $f \sim \langle E E \rangle/p \sim 1/\alpha_s \gg 1$ and has a finite albeit anisotropic momentum extent. As illustrated in the central panel of \fig \ref{fig:init_stages} at the example of $f(p_\perp, p_z{=}0)$, the plasma approaches a classical self-similar attractor far from equilibrium that exhibits universal dynamics $f(\tau, p_\perp, p_z) = \tau^\alpha f_s(\tau^\beta p_\perp, \tau^\gamma p_z)$. In \cite{Berges:2013eia, Berges:2013fga} the universal scaling exponents and distribution $f_s$ were extracted, and it was demonstrated that the scaling exponents agree with those predicted by the first stage of the `bottom-up' thermalization scenario of \cite{Baier:2000sb} that we will discuss below. Moreover, this universal regime shares a far-from-equilibrium universality class with scalar systems \cite{Berges:2014bba, Berges:2015ixa, PineiroOrioli:2015cpb} and even exhibits prescaling \cite{Mazeliauskas:2018yef, Brewer:2022vkq, Heller:2023mah}, where the distribution reaches its universal scaling form before the scaling exponents.

\subsubsection*{Bottom-up thermalization: QCD kinetic theory}

   When quasiparticles have formed, kinetic theory becomes applicable.%
   \footnote{Note that this assumes narrow excitations in gluon and fermion spectral functions. While for isotropic plasmas gluonic excitations are indeed narrow for all momentum modes \cite{Boguslavski:2018beu}, this may not be the case for fermionic excitations at the lowest momenta \cite{Boguslavski:2021kdd} and for gluons at low momenta and extremely anisotropic distributions $f(\vb p)$ \cite{Boguslavski:2021buh, Boguslavski:2019fsb}.}
   This happens at $\Q \tau \sim 1$, such that the classical attractor regime of the over-occupied plasma can already be studied using kinetic theory \cite{Baier:2000sb, Kurkela:2015qoa}. This is part of the kinetic bottom-up thermalization (or `hydrodynamization') scenario \cite{Baier:2000sb} that consists of three stages, as illustrated in the third and fourth panel of \fig \ref{fig:init_stages}: 1) the classical attractor as discussed above, 2) the momentum anisotropy freezes (represented by the pressure ratio $P_T/P_L$ on the $y$ axis in the fourth panel) while the typical hard occupancy decreases ($x$ axis of the fourth panel), 3) a stage of radiational breakup of hard modes, which leads to a momentum cascade toward the soft sector and eventual hydrodynamization. Time markers in the fourth panel roughly separate these stages, triangles at $P_T/P_L = 2$ mark a time close to equilibrium.

   Numerically, QCD effective kinetic theory (EKT) simulations can be used \cite{Arnold:2002zm, Kurkela:2015qoa, Kurkela:2018xxd, Du:2020zqg} that solve
   \be
    -\pdv[f({\vb p})]{\tau} = {\mathcal {C}^{1\leftrightarrow 2}}[f({\vb p})]+ {\mathcal{ C}^{2\leftrightarrow 2}} [f({\vb p})] - \frac{p_z}{\tau} \pdv[f({\vb p})]{p_z}
   \ee
   and describe the evolution of fermionic and gluonic distributions under the influence of Bjorken expansion using elastic $2 \leftrightarrow 2$ and inelastic effectively $2 \leftrightarrow 1$ scattering processes. The K{\o}MP{\o}ST framework \cite{Kurkela:2018wud} extends EKT simulations in a homogeneous plasma by providing energy-momentum tensor fluctuations for the subsequent hydrodynamic regime.
   Eventually, EKT simulations smoothly transition to hydrodynamics \cite{Kurkela:2015qoa} around one or few fm/c, as illustrated in the last panel of \fig \ref{fig:init_stages}.

\subsection{Further recent approaches to initial stages}

\noindent We will now highlight some recent developments and approaches to initial stages.

It has been argued recently \cite{Ambrus:2022qya} within a kinetic theory approach that the onset of hydrodynamics requires sufficiently large colliding nuclei like in Pb + Pb or Au + Au collisions while for smaller systems as for O + O collisions the applicability of hydrodynamics is uncertain. Thus, in this case, the pre-equilibrium regime of the QGP may be particularly important for observables. 

The hydrodynamic stage of the QGP can be extended to earlier times by identifying hydrodynamic attractors \cite{Heller:2015dha, Almaalol:2020rnu, Soloviev:2021lhs}. Such attractors allow establishing relations between initial state properties and final state observables more easily \cite{Giacalone:2019ldn} and reveal a rich structure of universal features in the dynamics \cite{Soloviev:2021lhs}. 
Recent studies also identify universal and attractor properties of the energy-momentum tensor components and correlations going beyond hydrodynamic modes \cite{Brewer:2022ifw, Du:2023bwi}.

At large gluon occupancies, a gauge-invariant condensate has been argued to emerge during the classical attractor regime \cite{Berges:2019oun, Berges:2023sbs}. The analysis is based on Wilson loop expectation values and on correlation functions of order parameters including the spatial Polyakov loop and a scalar field associated with the exponent of the Polyakov loop. With an effective theory formulation for this scalar field, a condensate at early times may impact some aspects of the initial stages.

Since available theoretical methods are limited to parts of the QGP evolution and rely on approximations, a general ab-initio method for the pre-equilibrium dynamics of QCD becomes necessary. Moreover, evolution equations for hard probes and hydrodynamic equations require the knowledge of transport coefficients. 
However, direct computations of such QCD observables using a real-time lattice formulation are currently not feasible due to the infamous sign problem in the partition function $Z = \int D A\,e^{i S[A]}$ for real-time paths.
The Complex Langevin (CL) method is a promising approach that may allow numerical simulations on complex time contours by solving CL equations for gauge theories, here written for a Yang-Mills theory in the continuum,
\begin{align}
    \frac{\partial A^a_\mu(\theta, x)}{\partial \theta} &= i \frac{\delta S_\mathrm{YM}}{\delta A^a_\mu(\theta, x)} + \eta^a_\mu(\theta, x), \qquad S_\mathrm{YM} = - \frac{1}{4} \int_{\mathscr C} d^4x F^{\mu \nu}_aF_{\mu \nu}^a,
\end{align}
with a Gaussian distributed field $\eta$ and $\mathscr C$ denoting the complex time contour. Observables are computed by averaging over sufficiently late fictitious Langevin times $\theta$. Previous studies of SU(2) theory in thermal equilibrium in 3+1 dimensions showed that usual CL simulations often converge to wrong values \cite{Berges:2006xc}. Exploiting the kernel freedom of CL, a new anisotropic kernel was introduced in \cite{Boguslavski:2022dee} that improves the stability of the stochastic Langevin process and leads to correctly converged results on complex time paths. By going to real times, the CL method with the new kernel has the potential to allow us to extract transport coefficients in QCD directly and eventually to study the non-equilibrium dynamics of QCD from first principles.


\section{Probing the pre-equilibrium medium evolution}
\label{sec:probe_pre_equ}

\subsection{Hard probes: a window into initial stages}

Hard probes have a high potential to provide signatures of the pre-equilibrium evolution of the QGP during initial stages. It is constructive to distinguish two forms of hard probes that are fundamentally different in their nature, origin, and interactions with the QGP. 

{\em Electromagnetic probes} include direct photons and dileptons (lepton-antilepton pairs), see \cite{Geurts:2022xmk} for a recent review. They are produced at all stages of the evolution and, due to their electromagnetic interactions, these hard probes generally do not interact with the medium once created. As they can originate also from the pre-equilibrium regime of the QGP, their yields, transverse momentum distributions, elliptic flow $v_2$, or nuclear modification factors $R_{AA}$ could contain signatures of the initial stages of the QGP.
For instance, it has been shown recently in \cite{Hauksson:2023dwh} that momentum anisotropies of quasiparticle distributions as typically encountered during the bottom-up thermalization scenario of the QGP can lead to a polarization of photons. In particular, direct photons emitted during these stages are dominantly polarized along the beam axis. 
In a different study \cite{Coquet:2021lca}, the production of dileptons during the pre-equilibrium kinetic stage is computed and compared to the dilepton production from the thermal QGP and from the Drell-Yan background. It was indicated that dilepton spectra with an invariant mass within $1 < M < 5$ GeV have the potential to provide valuable signals of the pre-equilibrium stages, with sensitivity to the momentum anisotropy and underoccupancy of quark distributions of the pre-equilibrium QGP. 

In contrast, jets, heavy quarks, and quarkonia are {\em QCD probes} and thus constantly interact with the QGP medium during their evolution (see \cite{Apolinario:2022vzg} for a recent review). Therefore, their final-state observables like spectra, elliptic flow, or $R_{AA}$, show cumulative effects from all the different stages of the medium evolution. Since this includes the pre-equilibrium regime, these hard probes also provide a window of opportunities to study the pre-equilibrium properties of the QGP.


\begin{figure}[t]
    \centering
    \quad\includegraphics[width=0.22\textwidth]{\pToFigs/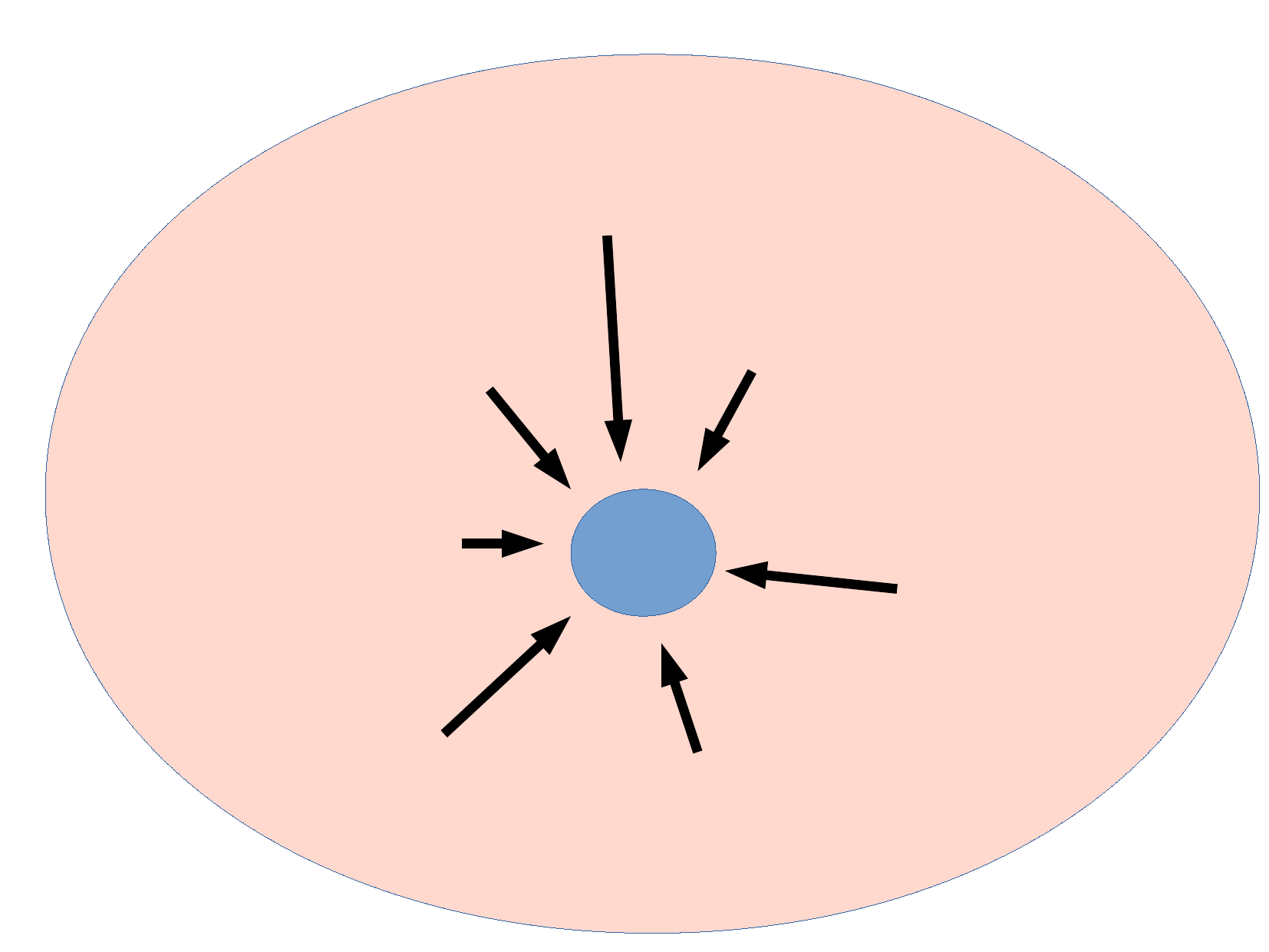}
    \qquad\qquad\qquad
    \includegraphics[width=0.32\textwidth]{\pToFigs/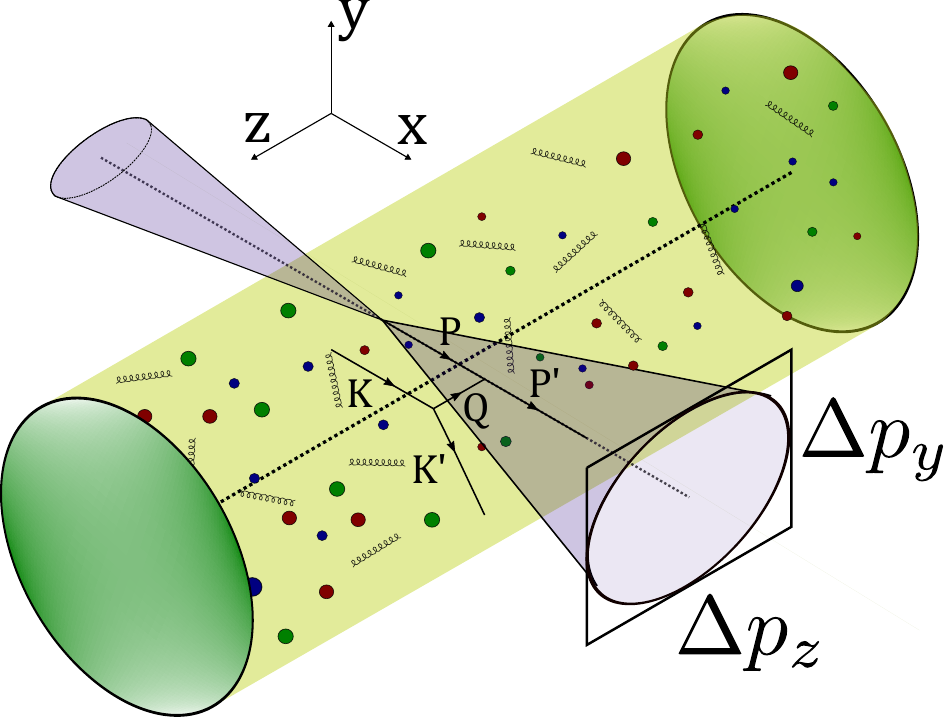}
    \caption{Simplified visualizations of {\em (left:)} heavy quark diffusion, and of {\em (right:)} jet quenching in the QGP via momentum broadening of the leading jet parton (in $x$ direction). (right figure taken from \cite{Boguslavski:2023xxx})}
    \label{fig:kappa_qhat_visuals}
\end{figure}

\begin{figure}[t]
    \centering
    \includegraphics[width=0.45\textwidth]{\pToFigs/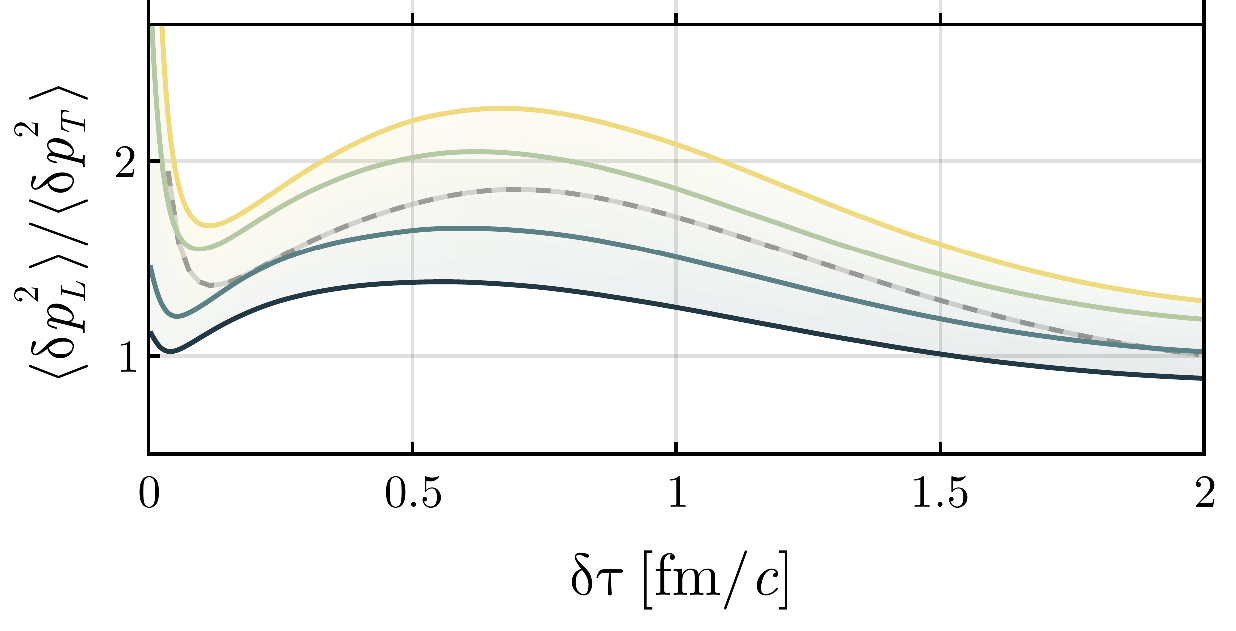}
    \qquad
    \includegraphics[width=0.475\textwidth]{\pToFigs/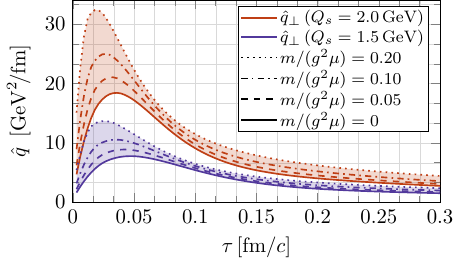}
    \caption{Transport coefficients during the Glasma stage. {\em (Left:)} approximately $\kappa_z / \kappa_T$ of beauty quarks, {\em (right:)} $\hat q$ of jets, as functions of time. (figures taken from \cite{Avramescu:2023qvv} and \cite{Ipp:2020nfu})}
    \label{fig:kappa_qhat_Glasma}
\end{figure}

\begin{figure}[t]
    \centering
    \includegraphics[width=0.45\textwidth]{\pToFigs/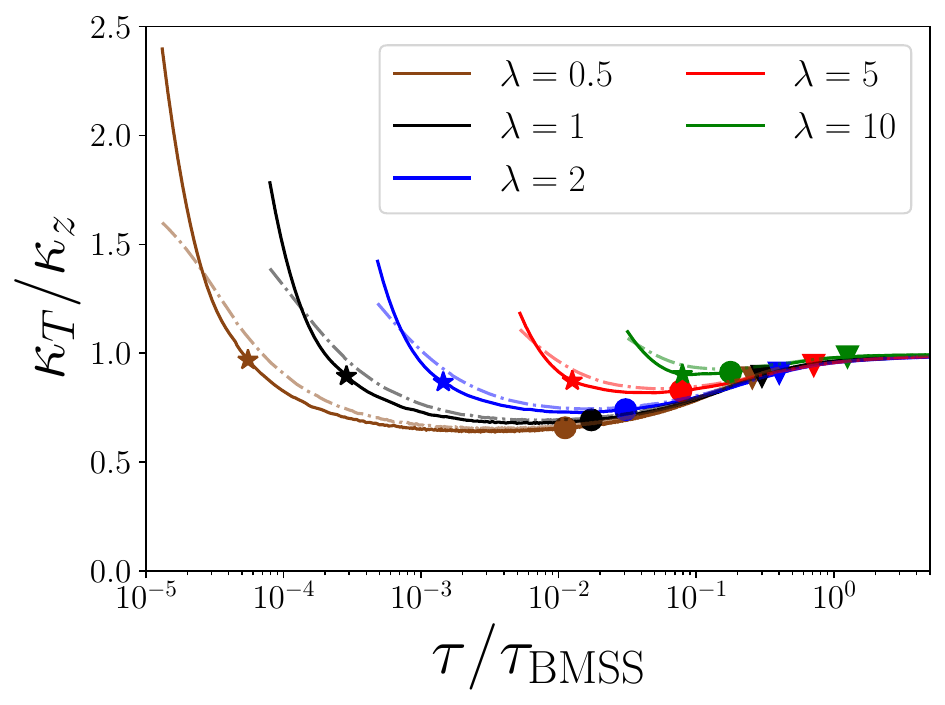}
    \qquad
    \includegraphics[width=0.45\textwidth]{\pToFigs/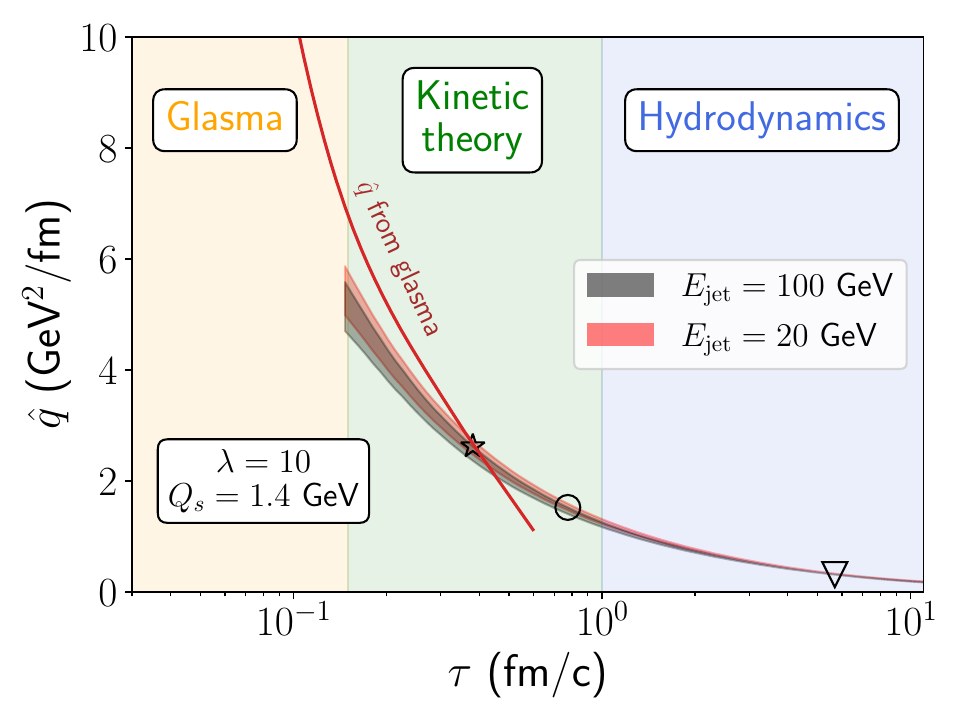}
    \caption{Transport coefficients during the kinetic bottom-up stages as functions of time. {\em (Left:)} $\kappa_T / \kappa_z$ of heavy quarks, {\em (right:)} $\hat q$ of jets -- each band includes simulation results for different initial conditions and time-dependent transverse momentum cutoff models. To guide the eye, the same time markers are added as in the fourth panel of \fig \ref{fig:init_stages}. (figures taken from \cite{Boguslavski:2023fdm} and \cite{Boguslavski:2023alu}, the Glasma line stems from \cite{Ipp:2020nfu})}
    \label{fig:kappa_qhat_EKT}
\end{figure}

\subsection{Transport coefficients $\kappa$ and $\hat q$ from the pre-equilibrium QGP}

The influence of the QGP on the evolution of QCD probes can be efficiently characterized by transport coefficients. These can be calculated within the QGP medium, and they enter effective evolution equations of the hard probes. 

Let us focus on the momentum broadening of heavy quarks with a large mass, $M \gg p, \Q$, and of jets, i.e., highly energetic partons with jet momentum $p \gg \Q$ that traverse the medium. These processes are visualized in \fig \ref{fig:kappa_qhat_visuals} in a simplistic manner. Once created, heavy quarks receive `kicks' from the QGP constituents, leading to momentum broadening which is parameterized by the heavy-quark diffusion coefficient $\kappa_i$ \cite{Moore:2004tg}, where we distinguish broadening into the beam direction $z$ (or $L$) and transverse to it denoted by $T$. Moreover, this parameter also enters the Lindblad evolution equation for quarkonia \cite{Brambilla:2016wgg, Brambilla:2020qwo}. Similarly, the jet quenching parameter $\hat q_i$ quantifies momentum broadening transverse to the jet's trajectory along the path $L$ due to elastic interactions with the QGP constituents via the collision kernel $C$, and is shown schematically in the right panel of \fig \ref{fig:kappa_qhat_visuals} for a jet in $x$ direction. In summary, these transport coefficients read 
\begin{align}
    \kappa_i = \frac{\ud}{\ud \tau}\langle p_i^2 \rangle\,, \qquad 
    \hat q_{i} = \frac{\ud}{\ud \tau}\langle p_{\perp,i}^2 \rangle = \int \ud^2 q_\perp\, q_{\perp, i}^2\, C(\vb q_\perp)\,,
\end{align}
with transverse momenta $q_{\perp, i}$ with $i = y$ or $z$.
It has been pointed out recently that pre-equilibrium dynamics may have sizable effects on heavy-quark diffusion and jet quenching \cite{Sun:2019fud, Ipp:2020nfu, Boguslavski:2020tqz, Khowal:2021zoo, Carrington:2021dvw, Andres:2022bql, Barata:2022utc}. 

More specifically, the transport coefficients $\kappa$ and $\hat q$ have been recently computed during the Glasma phase \cite{Ipp:2020nfu, Ipp:2020mjc, Boguslavski:2020tqz, Carrington:2020sww, Carrington:2021dvw, Carrington:2022bnv, Avramescu:2023qvv} using force-force correlations or, alternatively, Wong's equations in classical-statistical simulations. Their evolutions are shown in \fig \ref{fig:kappa_qhat_Glasma} for the ratio $\kappa_z / \kappa_T$ of $b$ quarks in the left panel and for $\hat q$ in the right panel. The resulting values turn out to be large; for $\hat q$ they exceed typical thermal values during the hydrodynamic evolution of the QGP by an order of magnitude. Moreover, the coefficients are anisotropic with $\kappa_z > \kappa_T$ and $\hat q_z > \hat q_y$ in the Glasma, and this ordering seems to originate mainly from an asymmetry between chromo-magnetic and -electric fields at these early times \cite{Ipp:2020nfu}. 

To bridge the gap between the Glasma and hydrodynamic regimes, new studies using QCD kinetic theory simulations have been conducted in \cite{Boguslavski:2023alu, Boguslavski:2023fdm} where $\hat q_i$ and $\kappa_i$ have been computed during the bottom-up hydrodynamization scenario (c.f., \se \ref{sec:init_stages_overview} and \fig \ref{fig:init_stages}). 
Most recent kinetic theory results also include the heavy quark drag and diffusion coefficients \cite{Du:2023izb}.
In \fig \ref{fig:kappa_qhat_EKT}, the pre-equilibrium evolution of the ratio $\kappa_T / \kappa_z$ is shown in the left panel (note the different ratio from $\kappa_z / \kappa_T$ as for the Glasma in \fig \ref{fig:kappa_qhat_Glasma}) and of $\hat q$ in the right panel. At the example of jet momentum broadening, one observes that $\hat q$ smoothly connects the large values of the Glasma with the lower ones in the hydrodynamic phase. Remarkably, the pre-hydrodynamic evolution of $\hat q$ during the bottom-up kinetic scenario shows little sensitivity to the initial conditions, jet energies, and models of the transverse momentum cutoff \cite{Boguslavski:2023alu}.%
\footnote{Since $\hat q$ requires a transverse momentum cutoff $\Lambda_\perp > q_\perp$ in the limit of large jet momentum, cutoff models are used that depend on fixed jet energy $E$ and time-dependent effective medium temperature $T$ from Landau matching. The bands in the right panel of \fig \ref{fig:kappa_qhat_EKT} contain curves with the LPM cutoff $\Lambda_\perp^{\text{LPM}} \sim (ET^3)^{1/4}$ and kinematic cutoff $\Lambda_\perp^{\text{kin}} \sim (ET)^{1/2}$.}
Furthermore, during most of the kinetic evolution, one finds the same ordering $\kappa_z > \kappa_T$ and $\hat q_z > \hat q_y$ as in the Glasma, albeit with a different origin: It is a consequence of the large momentum anisotropies and low occupancies of the distribution in the second and third stages of the bottom-up scenario. For $\hat q$ the anisotropic broadening between $\hat q_z$ and $\hat q_y$ reaches up to $20 \%$ in the kinetic regime \cite{Boguslavski:2023alu} for the physical parameters used in \fig \ref{fig:kappa_qhat_EKT}.


\section{Conclusion}
\label{sec:concl}

We have given a short overview of the initial stages of the QGP in heavy-ion collisions with particular emphasis on the different phases, their methodological approaches, and new developments. Focusing on recent examples, we have also discussed exciting opportunities that hard probes may provide to test our theoretical understanding of the initial stages. 

In particular, based on weak-coupling methods, our current picture of the pre-equilibrium QGP dynamics involves the Glasma consisting of large classical fields that triggers plasma instabilities, leading to a highly occupied plasma of quasiparticle excitations. At this point, the classical-statistical theory description can be switched to QCD kinetic theory simulations, which describe the three stages of the bottom-up thermalization scenario. Toward the end of its pre-equilibrium evolution, the QGP smoothly transitions to a hydrodynamic description. 

We have discussed recent approaches and developments of this picture that include system size dependence for the applicability of hydrodynamics, hydrodynamic attractors and universal properties, and the emergence of gauge-invariant condensates. With a new technical advancement, the Complex Langevin method for gauge theories may become a powerful tool to extract transport coefficients directly and to provide a general ab-initio approach for the pre-equilibrium QGP.

Experimentally, hard probes can provide the means to test our theoretical understanding of the initial stages. More specifically, recent studies on electromagnetic probes suggest that the polarization of direct photons and the invariant-mass spectra of dileptons may provide signatures of the pre-equilibrium QGP. Recent Glasma and kinetic theory studies of the transport coefficients $\hat q$ and $\kappa$ for momentum broadening of jets and heavy quarks also show very promising results. They indicate that both $\hat q$ and $\kappa$ are large during the initial stages of the QGP evolution and should not be neglected in phenomenological models of jet quenching, heavy-quark diffusion, and quarkonium evolution. 
Their anisotropic momentum broadening as $\hat q_z > \hat q_y$ extends over different initial stages in the pre-equilibrium QGP and could have observable consequences. For instance, if the medium anisotropy is sizeable as the jet escapes the medium, it could lead to a polarization of jet partons \cite{Hauksson:2023tze}. This condition may be fulfilled in medium-sized systems like O + O collisions if the medium is out of equilibrium at that time. However, further studies may be needed to quantify such effects.


\bibliographystyle{JHEP_mod}
\bibliography{HP2023}

\end{document}